# Analyzing the Dual-Path Peer-to-Peer Anonymous Approach

Ehsan Saboori
K.N Toosi University of Technology
Tehran, Iran

Majid Rafigh
Shahid Beheshti University (Former National University)
Tehran, Iran

Alireza Nooriyan
Iran University of Science and Technology
Tehran, Iran

## ABSTRACT
Dual-Path is an anonymous peer-to-peer approach which provides requester anonymity.This approach provides anonymity between a requester and a provider in peer-to-peer networks with trusted servers called suppernode so the provider will not be able to identify the requester and no other peers can identify the two communicating parties with certainty.Dual-Path establishes two paths for transmitting data. These paths called Request path and Response path. The first one is used for requesting data and the second one is used for sending the requested data to the requester. As Dual-Path approach is similar to Crowds approach, this article compares reliability and performance of Dual-Path and Crowds. For this purpose a simulator is developed and several scenarios are defined to compare Dual-Path and Crowds in different situations.In chapter 2 and 3 Dual-Path and Crowds approachesarebriefly described. Chapter 4 is talking about simulator. Chapter 5 explains the scenarios for comparison of performance. Chapter 6 is about comparison of reliability and chapter 7 is conclusion.

## General Terms
Network Security, Peer-to-Peer Networks,Anonymous network,Dual-Path Anonymous communication

## Keywords
Dual-Path, Peer-to-peer Networks, Anonymity, Onion Routing, Crowds,

## 1. INTRODUCTION
A peer-to-peer network is a dynamic and scalable set of computers also referred as peers or nodes. The peers can join or leave the network at any time. [1]These nodes can connect to each other and they can send or receive data. One of the most important issues in this kind of networks is privacy. It means a node can participate in the network in such a manner that nobody can compromise its activity during its participation.One aspect of privacy is anonymity. Anonymity means nobody can compromise the identity of a specific node in the network when it sends or receives data or does other activities.Dual-Path is an anonymous peer-to-peer approach which provides requester anonymity. [2] There are some peer-to-peer anonymous approaches that provide anonymity in this kind of networks. Crowds [3], Hordes [4], Freenet [5], Tor [6] and Tarzan [7] are proposed anonymous approaches. Crowds is an anonymous web transaction protocol and one of the oldest anonymizer networks and only provides requester anonymity. Crowds contains a closed group of participating nodes called jondos and uses a trusted third party as a centralized crowd membership server called blender. [3] There are three kinds of anonymity: Provider anonymity that hides the identity of a provider against other peers, Requester anonymity that hides a requester's identity and Mutual anonymity that hides both provider's and requester's identities. In the most stringent version, achieving mutual anonymity requires that neither the requester, nor the provider can identify each other, and no other peers can identify the two communicating parties with certainty. [8] Dual-Path approach provides requester anonymity to protect the identity of the requester and the transferred data against other peers specially the intruders. It is based on Onion Routing [10] mechanism. Onion Routing is the technique in which the requester and the provider communicate with each other anonymously by means of some intermediate peers called as onion routers. In this technique, messages route between onion routers. The messages encrypted with onion router's public key. Each onion router learns only the identity of the next onion router. [10]

## 2. CROWDS
Crowdsis an anonymous web transaction protocol and one of the oldest anonymizer networks and only provides requester anonymity. Crowds contains a closed group of participating nodes called jondos and uses a trusted third party as centralized crowd membership server called blender. The new jondo requests crowd membership from the blender, then the blender replies with a list of all current crowd members. After that, the blender informs all previous members of the new member. The requester node selects randomly a jondo from the member list and forwards the request to it. The following nodes decide randomly whether to forward the request to another node or to send it to the server. Crowds is vulnerable to DoS attacks.[3] Figure 1 shows the concept of Crowds approach.

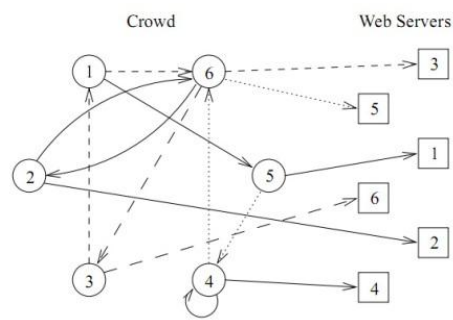

**Fig 1: Crowds Concept [3]**

## 3. DUAL-PATH
The basic principle of this approach is to relay messages from requester to provider through multiple intermediate peers so that the true origin and destination of the messages is hidden from other peers. The requester creates a dual-path which contains a path to send request and another to get respond from provider so that the provider cannot compromise the requester's identity. The transferred data between requester and provider is encrypted to protect it against eavesdropping. So, in this approach there are two paths to connect requester to provider: request path and response path. Both of them are initiated by requester randomly. The requester can change these paths randomly while connecting to provider at any time. [2]Figure2 illustrates a request and response paths in the network.

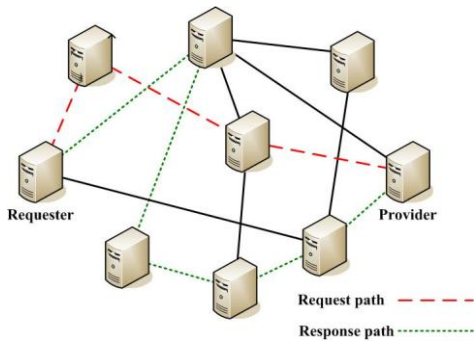

**Fig 2: Dual-Path paradigm (The base principle of the request and response paths)[2]**

To create paths, a requester requests a list of all current peers. After that the requester can choose two sets of peers randomly or depending on traffic information or other parameters. The response path is embedded in request path by requester and when the provider gets a message from the requester it knows which peer must be given the response message as the first. For this purpose, the requester creates a wrapped message and sends it to the next peer. The next peer decrypts the message and sends it to the next peer which is determined in the message. This process is continuing until the message is being received by the provider. When the provider wants to response the request, it sends the respond message to the peer that determines in the tail of the received message from the requester. The next peer does the same action until the massage is being received by the requester.[2]Lets consider peers P1, P2 and P3 which are chosen randomly by requester for request path and P4, P5 and P6 which are chosen for response path. Also consider M, the message, which the requester wants to send. Figure3 shows the Dual-path created by requester. In this figure, "A" acts as a requester and "B" acts as a provider. "A" creates two paths to communicate with "B" and sends messages via them. "A" must rely messages through P1, P2 and P3 (request path) to send them to provider. Also "A" receives the response of its request through P4, P5 and P6 (response path).[2]

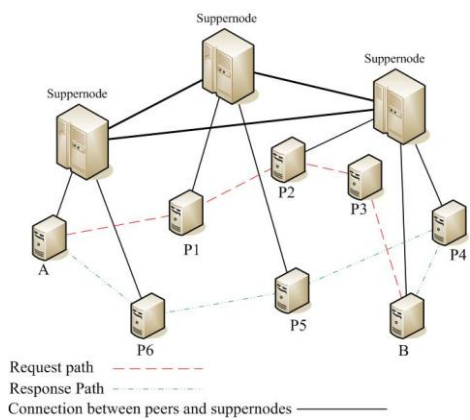

**Fig 3: Dual-Path between A and B as requester and provider[2]**

After the requester (A) creates the Dual-Path, now it must create the packet of the messages. To create the packets, the requester (A) must encrypt the messages by intermediate peers' public keys in a layer by layer structure, such as onion routing [10] mechanism. Figure 4 shows how the requester wraps the message by intermediate peers public keys.[2]

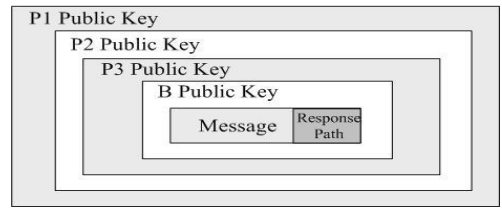

**Fig 4: Use intermediate peers' public keys to wrap message[2]**

While the requester wraps the message, it embeds the response path in the end of message as it is shown in Figure4. This part of message contains the response path. The structure of response path in wrapped message is illustrated in Figure5.[2]

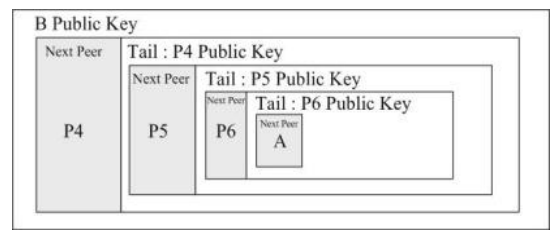

**Fig5: The structure of response path in wrapped message[2]**

When the provider (B) receives the packets, it extracts the message and the response path packet. Each response packet has two parts, the "Next Peer" and the "Tail". The "Next Peer" part contains the next peer in which the message must be sent to it. After extracting the response path packet by provider (B), it encrypts the response message by P4 public key and attaches the "Tail" part of response path packet at the end of it. Now the provider (B) sends the wrapped message to P4. P4 does the same process and sends the received messages to P5. P5 sends the messages to P6 and at last, P6 sends the messages to the requester (A). When the requester received its response, one dual-path cycle is completed and the requester can use this dual-path or choose another dual-path for more security in order to transfer the messages.[2]

## 3.1 Dual-path advantages

Dual-Path uses several intermediate peers to connect requester to provider and change response/request path randomly. Therefore this approach increases reliability, because if one intermediate peer suddenly leaves the network, the requester will choose another path to connect to provider. [2]Also it can consider the network traffic to use the paths with less traffic in order to increase efficiency and reduce the delay time. Dual-Path has a good performance when the network traffic is increased. One of the most important advantages of Dual-Path is more prevention against traffic analysis, because the requester can change dual-path periodically so it is too difficult for the intruder to reveal the origin path of transferred data. The intruder cannot use time-to-live attack against network, because each path has different time to live so the intruder cannot gather useful information to reveal requester.[2]Time-to-live counters determine the maximum number of hops for a message and are used in most peer-to-peer networks to avoid flooding. If an attacker can send a request to a node with such a low time-to-live counter that the packet will probably not be forwarded, any response relieves that note as the provider. [10]

## 4. SIMULATION

A simulator needed to compare performance and reliability of Crowds and Dual-Path. For this purpose the simulator developed by C# .Net. The server configuration is Intel core i7 1.6GHz, 4.00 GB Ram and Windows 7 64Bit.For the simplicity of the simulation supposed that there is only one suppernode and all nodes are in one group. This assumption does not effect on the Dual-path approach, because the goal is to send data from provider to requester.There are several parameters considered in simulation.Delay time for sending TCP packets. This delay is 1.92ms for sending 1KB data between two nodes with 128Kbps connection. Encryption time for 1KB data with RSA algorithm and with 1024bit key is 0.8ms. Decryption time for 1KB data with RSA algorithm and with 1024bit key is 9.3ms. The size of data varies between 1 to 100KB. The number of nodes varies between10 to 1000 nodes. The dropping ratio for all nodes is 40%. It means the possibility that each node maybe out of service is 40% of the time. The network traffic is considered as both static and dynamic. The decision ratio is 50%. For Crowds approach it means the probability of decision to send packets to the destination or send them to other node.For Dual-path approach it means the probability of changing the paths.

## 5. PERFORMANCE

In this chapter performance of Dual-Path is compared to performance of Crowds.The delay time for sending data from provider to requester is measured for both approaches. This time is calculated in specific period of time. There are several scenarios. In each of them, one or more parameters are changed. These parameters are size of sending data, network traffic, and number of nodes.For example in a scenario, only the size of sending data is increased and simulator log the delay time for sending 1KB, 2KB,..., 100KB for Dual-path and Crowds approach. After that the log data are plotted in chart. In all charts the vertical axis shows the delay times and the horizontal axis shows the times. So these charts illustrate the behavior of these approaches toward changing the parameters.For having a more reliable value for delay time, sending data is repeated 5000 times and the average of these valuesare considered as the delay time, because Crowds and Dual-path are based on probability. For instant, when simulator sendsa packet with 1KB size, it may takes 1ms and for other iteration it takes 1.1ms because the path is selected randomly and the delay time for each path is different.

### 5.1 Static parameters

In this scenario all parameters are static and will not change. So in this scenario the number of nodes, network traffic and size of sending data are the same in both approaches. Figure 6 illustrates the delay time for both approaches. As it shows, the delay time for transferring data between provider and requester in Crowds approach is more than Dual-path approach.

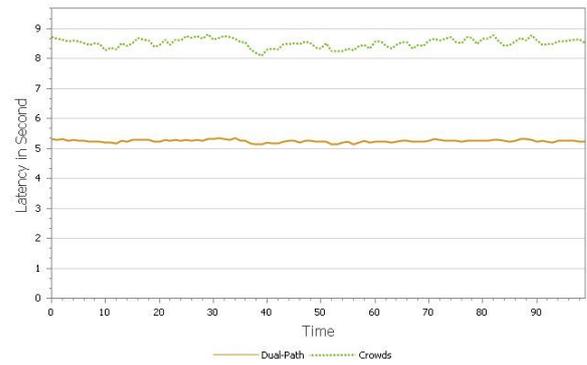

**Fig 6: Delay time for static parameters**

To have a better understanding, the improvement ratio is calculated for all scenarios.The improvement ratio is the differentiation of the delay time of Crowds in time *i* and the delay time of Dual-path is divided by 100.

$$\text{Improvement Ratio} = \frac{\text{Crowds}_{D_i} - \text{DualPath}_{D_i}}{100}$$

Where $D_i$ is delay time for each approach in time *i*.

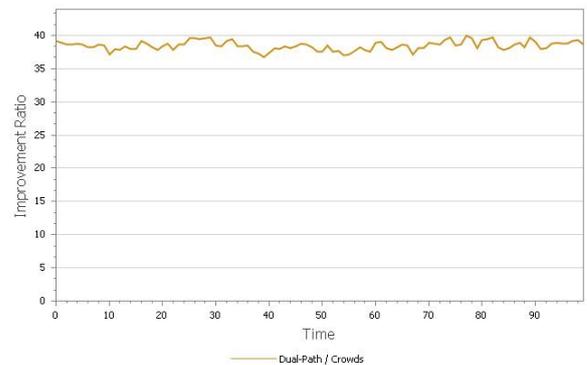

**Fig 7: Improvement ratio for static parameters**

Figure 7 shows the improvement ratio for this scenario. As it shows, Dual-path is approximately 37.5% faster than Crowds.

### 5.2 Increasing network traffic

In this scenario network traffic changesthrough the time and other parameter are static. For increasing the network traffic, simulator inserts delay in sending data between nodes and increases this delay over the time. Figure 8 illustrates the delay time for Crowds and Dual-path approaches. As it shows, the delay time in Crowds increases rapidly by increasing network traffic. However the delay time for Dual-path increases slightly. So Dual-path approach is more resistant against increasing network traffic and it has a better performance in high traffic networks. As Dual-path approach can choose its paths, it can choose the paths from low network traffic, so the network traffic has low impact on Dual-path approach. The behaviors of these approaches can be modeled asequation:

$$\text{Dualpath behavior} = 0.0464X + 5.2110$$

$$\text{Crowds behavior} = -0.0002X^2 + 0.1614X + 7.0096$$

The behavior ofDual-Path approach is O (n) and the behavior of Crowds is O ($n^2$).

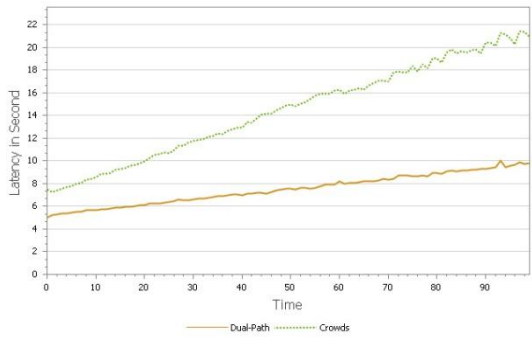

**Fig 8: Delay time for increasing network traffic**

Figure 9 shows the improvement ratio for this scenario. This trend shows that Dual-Path improvement ratio increases by increasing the network traffic. Eventually it has approximately 52.5% faster than Crowds.

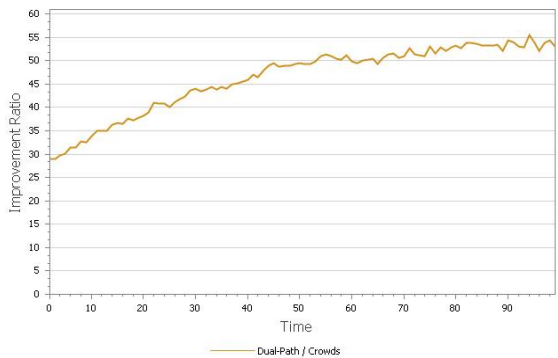

**Fig 9: Improvement ratio for increasing network traffic**

## 5.3 Increasing size of sending data

In this scenario size of sending data changesoverthe time and other parameter are static.The size of sending data increases from 1KB to 100KB. Obviously increasing the size of sending data increases the delay time. Figure 10 shows the behavior of two approaches in increasing the size of sending data. The delay time for Crowds is increased faster than Dual-path by increasing the size of sending data.

$$\text{Dualpath behavior} = -0.0002X^2 + 0.0757X + 0.0661$$

$$\text{Crowds behavior} = -0.0003X^2 + 0.1064X + 0.5352$$

The behavior of Dual-Path and Crowds is O ($n^2$). But the coefficient of $x^2$ in Dual-Path is smaller than Crowds. So Crowds is more sensitive to increasing the size of sending data.

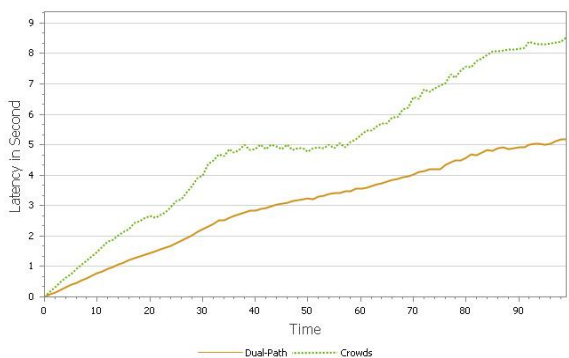

**Fig 10: Delay time for increasing the size of sending data**

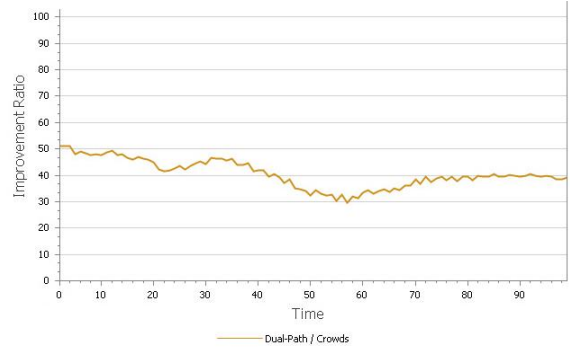

**Fig 11: Improvement ratio for increasing the size of sending data**

Figure 11 shows the improvement ratio for both approaches. It shows that the Dual-Path is about 30% to 50% faster than Crowds for increasing the size of sending data.

## 5.4 Increasing size of sending data and network traffic

In this scenario the size of sending data and network traffic are increased over the time. Figure 12 shows how the delay time is changed over the time in two approaches. The delay time rises rapidly in Crowds but it rises slightly in Dual-Path.

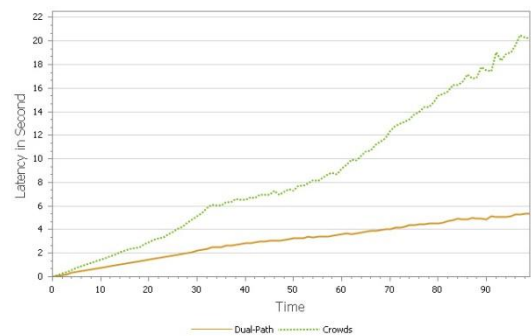

**Fig 12: Delay time for increasing size of sending data and network traffic**

Figure 13 shows the improvement ratio. It shows that the Dual-Path is faster about 45% to 75%.

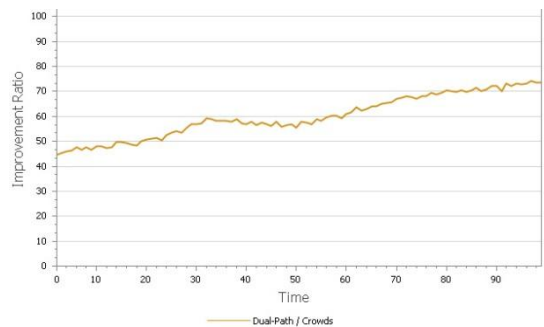

**Fig 13: Improvement ratio for increasing size of sending data and network traffic**

## 5.5 Increasing number of nodes

In this scenario the number of nodes changes through the time from 10 to 1000 and the other parameters are static for both approaches.Node means the peer or user (computer) that connected to the network. Figure 14 shows the delay time for

19

this scenario. Changing the number of the nodes does not have strong impact on the delay time. However Crowds is more sensitive rather than Dual-Path to increasing the number of nodes. Because Dual-Path selects two paths through nodes and changing the number of nodes is not important but Crowds should select a random node in each step. Dual-path is more stable in networks with dynamic nodes number.

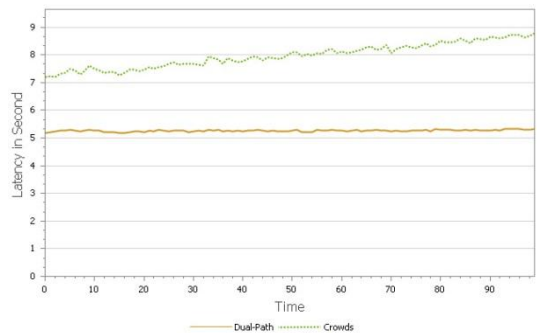

**Fig 14: Delay time for increasing the number of nodes**

Figure 15 shows the improvement ratio in this scenario. As this figure illustrates, Dual-Path approach is about 25% to 40% faster than crowds in changing the number of nodes.

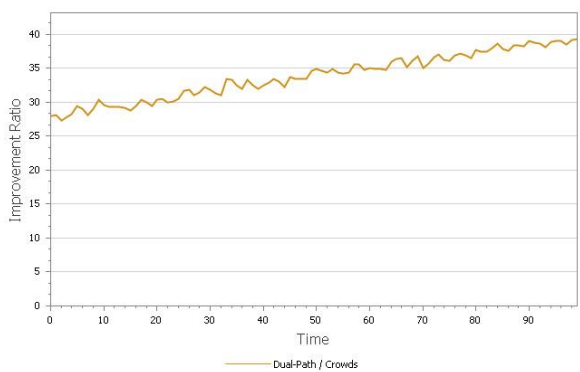

**Fig 15: Improvement ratio for increasing the number of nodes**

## 6. RELIABILITY

The other important factor to compare these two approaches is reliability. Reliability means to be able to expect that the data will be delivered to the requester properly. The most important issue in reliability is node failure. It means a node shutdown or stops working. If this node is used to relay data from provider to requester, the connection will be disconnected and the system now will be out of service. The simulator drop nodes randomly by 40% ratio in each iteration. Also the network traffic and the size of sending data change in some scenarios.

### 6.1 Increasing the node failure ratio

In this scenario 40% of nodes are dropped in each iteration and the other parameters are static. Figure 16 shows the delay time for both approaches. As it shows Crowds has more reliability than Dual-Path in nodes failure. Because in each step Crowds select a node randomly, so always there is an available node to relay the data. But in Dual-Path approach when a node is dropped and is used in paths, Dual-Path have to establish a new path to reconnect, so it needs more time and the delay time increases.

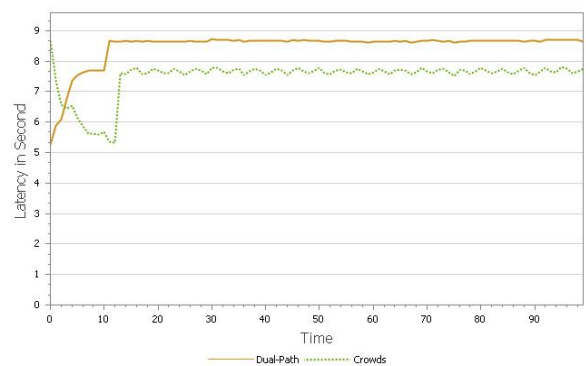

**Fig 16: Delay time for increasing nodes failure**

Figure 17 shows the improvement ratio for Dual-Path. Crowds is faster than Dual-Path when nodes are dropped in network and it has more performance in this situation about 12%.

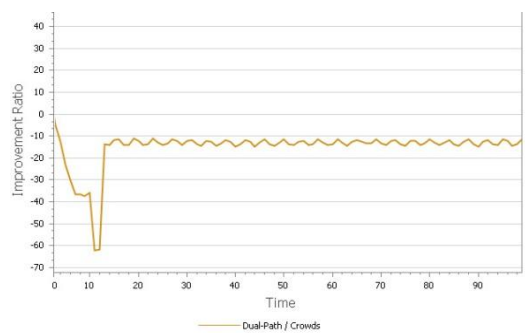

**Fig 17: Improvement ratio for nodes failure**

### 6.2 Increasing node failure and network traffic

In this scenario the nodes failure ratio and network traffic increase over the time. Figure 18 shows the delay time. The delay time in both approaches increases when the network traffic rises.

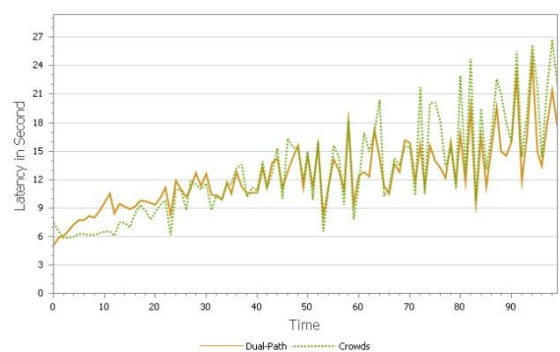

**Fig 18: Delay time for increasing nodes failure and network traffic**

Figure 19 shows the improvement ratio. As it shows, Dual-Path acts better when network traffic increases and nodes are dropped.

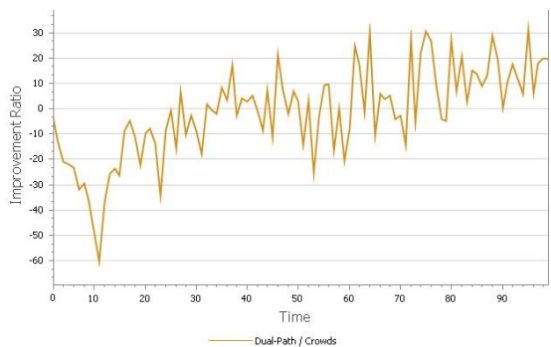

**Fig 19: Improvement ratio for increasing nodes failure and network traffic**

## 6.3 Increasing nodes failure ratio and size of sending data

In this scenario the size of sending data and nodes failure ratio rise over the time. Figure 20 shows the delay time for both approaches in this scenario. As it shows, the delay time increases with a liner behavior. But Crowds acts better.

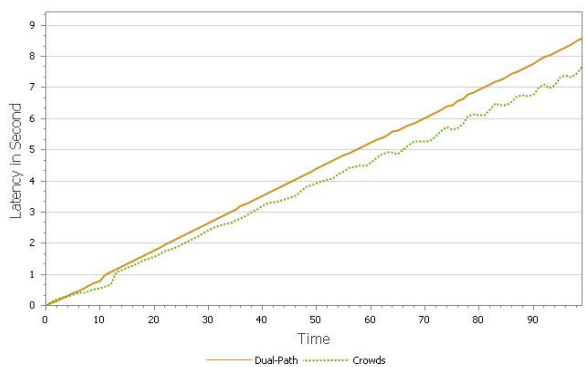

**Fig 20: Delay time for increasing nodes failure and size of sending data**

Figure 21 illustrates the improvement ratio. In this scenario, Dual-Path approach increases the delay time, However Crowds has a better performance about 10%.

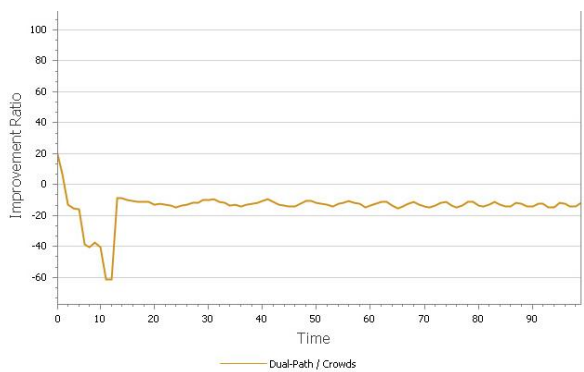

**Fig 21: Improvement ratio for increasing nodes failure and size of sending data**

## 6.4 Increasing nodes failure ratio, size of sending data and network traffic

In this scenario network traffic, size of sending data and nodes failure ratio are increased over the time. This scenario is the worst case for these approaches. Figure 22 shows the result.

The delay time is increased over the time for both approaches. But it's very various for Crowds and Dual-Path. They have a similar delay time. However the Crowds has a better delay time at first but by increasing network traffic and size of sending data Dual-Path will act better.

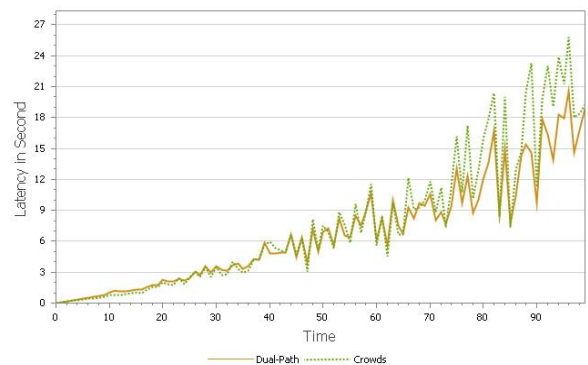

**Fig 22: Delay time for increasing nodes failure, size of sending data and network traffic**

Figure 23 illustrates the improvement ratio. As it shows, it's very various, but it can be concluded that by increasing the size of sending data and network traffic, the performance of Dual-Path increases.

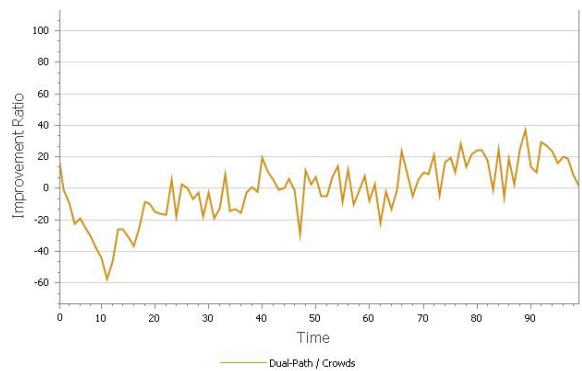

**Fig 23: Improvement ratio for increasing nodes failure, size of sending data and network traffic**

## 7. CONCLUSION

To compare performance of Dual-Path and Crowds, the delay time for sending data from provider to requester is measured in several scenarios. Each scenario studies about impact of one or more parameters on the delay time in both approaches. These parameters are Network Traffic, Number of Nodes and Size of sending data. Figure 24 illustrates the improvement ratio in all scenarios. The ratio1 is improvement ratio for the scenario that all parameters are static. The ratio2 shows improvement ratio in the scenario that network traffic and size of sending data increase. The ratio3 is improvement ratio in the scenario in which size of sending data increases. The ratio4 shows the condition that the network traffic increases over the time and the ratio5 shows the behavior of the improvement ratio when the number of the nodes increases.

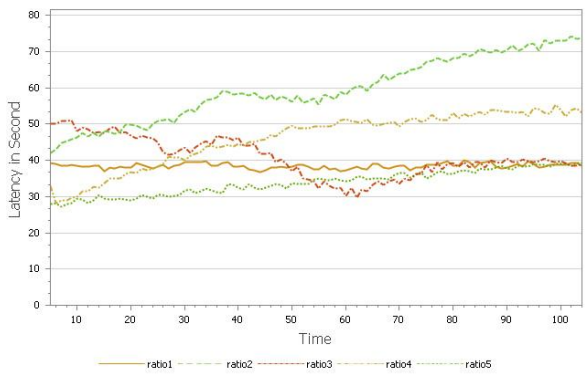

**Fig 24: Improvement ratio for all scenarios at a glance**

Figure 24 shows that the minimum improvement ratio is about 30% when the size of sending data and the network traffic increase together. The maximum improvement ratio is about 75% when network traffic and size of sending dataare increased in the network. To compare reliability, the delay time is measured while nodes are dropped randomly. There are three scenarios for comparison of reliability. Nodes failure while: 1- increasing size of sending data, 2- increasing network traffic and 3- increasing size of sending data and network traffic. Crowds is more reliable in first scenario but in the second and third scenarios, Dual-Path's reliability is increased over the time.Dual-Path has a better performance in all scenarios but Crowds is more reliable. However, when network traffic increases over the time Dual-Path acts better in both performance and reliability factors.